\documentstyle[12pt]{article}
\begin{document}
\baselineskip 18pt
    \begin{flushleft}
    January 12, 1994
    \hfill {IPSAS-RA-9401}
    \end{flushleft}
    \vspace{0.5cm}

\begin{center}

{\Large \bf

              RELATIVISTIC QUARK MODEL BASED DESCRIPTION
                     OF LOW ENERGY NN SCATTERING

}
\vspace{\baselineskip}
\vspace{\baselineskip}
 {\bf  R. Antalik }
              \footnote {E-mail address: antalik@savba.sk}  and
 {\bf  V.E. Lyubovitskij }
              \footnote {Permanent address: Department of Physics,
                      Tomsk State University, 634055 Tomsk, Russia}\\
\vspace{\baselineskip}
{\em
                         Institute of Physics,
                      Slovak Academy of Sciences
\\
                          842 28 Bratislava,
                               Slovakia
\\
}
\end{center}
\vspace{\baselineskip}
\vspace{\baselineskip}
\begin{abstract}
\noindent
A description of the NN scattering is constructed starting from the
$\pi, \eta, \eta'$ pseudoscalar-, the $\rho, \phi, \omega$ vector-,
and the $\varepsilon(600), a_0, f_0(1400)$ scalar - meson-nucleon
coupling constants, which we obtain within a relativistic quark model.
Working within the Blankenbecler-Sugar-Logunov-Tavkhelidze
quasipotential dynamics we thus describe the NN phase shifts in a
relativistically invariant way.  In this procedure we use the
phenomenological form factor cutoff masses and the effective
$\varepsilon$ and $\omega$ meson-nucleon coupling constants, only.
The comparison of our NN phase shifts to the both empirical data and
the Bonn OBEP fit shows good agreement -- the ratio of the $\chi^2$
for the present results to the $\chi^2$ for Bonn OBEP description is
1.2.  
\end{abstract}
\vfill
\section{Introduction}
\noindent
     One of main problems in subnuclear physics concerns the structure
and interactions of hadrons in terms of their elementary quark and
gluon constituents.  However, at low energies and small momenta, the
traditional description of nuclear forces and nuclear dynamics based
on nucleon and meson degrees of freedom appear to give a viable
phenomenology of the nuclear reactions and structure \cite{bj,ew}.

     Phenomenologically are nuclear forces understood in terms of
meson exchanges.  Their long range component, for the first time
introduced by Yukawa \cite{yukawa}, is generated by a pion exchange.
The intermediate range attraction between two-nucleons can be
understood in terms of a correlated two pion exchange, usually
simulated by a scalar-isoscalar $\varepsilon$ meson.  A repulsive
$\omega$ meson exchange represents the short range component of
two-nucleon forces and the $\rho$ exchange is notably distinctive in
the isovector-tensor channel.  In such one-boson exchange (OBE) model
approach, meson-nucleon coupling constants and form factor cutoffs
represent physical parameters, which are to be determined from the
best description of the NN scattering data.  The conventional
relativistically invariant interpretation of the experimental NN phase
shifts use such model of the NN forces, see Refs.
\cite{bj,nij78,paris,mhe,machleidt}.

     The microscopic interpretation of the NN scattering data has
mostly rested upon the various nonrelativistic quark models
\cite{okay,myhrer,shimizu}.  To preserve the relativistic invariance
of the microscopic interpretation, however, one has another
possibility, to use the QCD motivated, quantum field theory based
models, see, e.g., Refs.  \cite{namyslowski,mitra}.

     In this paper we investigate a possibility to understand the NN
scattering problem starting from a relativistically invariant quark
confinement model (QCM) developed at Dubna by Efimov, Ivanov and one
of us (V.L.) \cite{eiqcm,eiecaja,eildiquark,eiltomsk}.  Constructing
an OBE model of nuclear forces, we use our predictions of
meson-nucleon coupling constants that we obtain within the QCM.  The
form factors we treat but as is usual in a conventional interpretation
scheme \cite{bj,machleidt}, i.e., we choose their cutoffs to fit the
NN phase shifts.  This is justified because of the medium
renormalization of the meson and nucleon parameters, see, e.g.  Ref.
\cite{lurie}.

     A part of our interpretation scenario -- the QCM represents the
relativistically invariant effective quantum field theory variant
inferred from QCD.  Within this scheme are hadrons treated as composed
of quarks.  The confinement of quarks emerges as in the QCD through
nonperturbative gluon vacuum fields.  There is no attempt in this
model, however, to evaluate the quark confinement, but the S-matrix
integration measure itself is conveniently parameterized.  This
parameterization then allows us to evaluate all quark diagrams
representing the meson nucleon interactions.

     The processes investigated within this model approach cover the
static hadron characteristics, the strong, electromagnetic, and weak
dynamical properties of nonflavored, charmed, and bottom mesons and
baryons \cite{eiecaja,eildiquark,eiltomsk,qcm,al}.  In all these
studies acceptable results have been obtained.  Hence, it shows that
the physical picture behind the QCM represents the bulk properties
ofthe hadronic structure, although in the parameterized, nevertheless
in the unique way.

     In section 2, we will briefly specify the quasipotential dynamics
and the meson exchange model of the NN interaction we used here.  In
section 3, we will briefly review the meson-nucleon coupling constants
calculation within QCM.  In section 4, we present our results,
compare, and discuss them.  Section 5 is a conclusion.

\section{ NN scattering model }
\noindent
     To describe the scattering process we work in the framework of
the three dimensional quasipotential dynamics using the
Blankenbecler-Sugar-Logunov-Tavkhelidze equation \cite{bj}.  This
equation can be written for the R-matrix, which is directly related to
the NN phase shifts \cite{eah}.


     It is widely accepted \cite{bj,machleidt} that conventional
one-boson exchange model of the NN forces is capable to describe the
scattering observables.  The NN forces are then given as a sum of the
contributions of relevant mesons.  As the empirical findings show to
describe the low energy NN scattering the pseudoscalar, vector, and
scalar meson fields should necessary to be accounted for \cite{bj}.

     In the field theoretical language are meson-nucleon couplings
described by the following relativistically invariant Lagrangians for
pseudoscalar $\phi^{(ps)}$, scalar $\phi^{(s)}$ and vector
$\phi^{(v)}$ meson interactions

\begin{eqnarray}
{\cal L}_{ps}  = &i& \sqrt{4 \pi} \; g_{ps}\; \bar{\psi}
                     \gamma^{5}\psi \phi^{(ps)}  \; , \\
{\cal L}_{s}   = &i& \sqrt{4 \pi} \; g_{s}\; \bar{\psi} \psi
                     \phi^{(s)} \; , \\
{\cal L}_{v}   = &i& \sqrt{4 \pi} \; g_{v}\; \bar{\psi} \gamma_{\mu}
                     \psi \phi^{(v)}_{\mu}
                     \nonumber \\
               + &i& \sqrt{4 \pi} \; {f_{v} \over 4 M}\; \bar{\psi}
                     \sigma^{\mu \nu}
                     \psi ( \partial_{\mu} \phi^{(v)}_{\nu}
                     - \partial_{\nu}\phi_{\mu}^{(v)})  \; ,
\end{eqnarray}
     where $g$ and $f$ describe the vector and tensor couplings
of the nucleon field $\psi$ with $\alpha$ meson field
$\phi^{(\alpha)}$.

     Using Feynman techniques one can obtain the one-boson exchange
amplitudes for a particular mesonic field.  The pseudoscalar, scalar,
and vector meson amplitudes that we need for evaluation of the
Blankenbecler-Sugar-Logunov-Tavkhelidze equation in its R-matrix form
are explicitly shown in Ref.  \cite{mhe}.  The form factors applied at
each vertex are taken as

\begin{eqnarray}
F_{\alpha}(\Delta^{2}) =
  {\left({ \Lambda_{\alpha}^2 - m_{\alpha}^2
                            \over
           \Lambda_{\alpha}^2 - \Delta^{2}}\right)} \;,
\label{ff}
\end{eqnarray}
 where $\Lambda_{\alpha}$ is the cutoff of the ${\alpha}$ meson
 of a mass $m_{\alpha}$, and
$
\Delta^{2} = (E_{q'}-E_{q})^{2} - ({\bf{q'-q}})^{2}
$
is the four-momentum of the exchanged particle \cite{eah}.

\section{Meson-Nucleon Coupling Constant Calculations}
\subsection{Quark Confinement Model}
\noindent
     As the quark confinement model had been developed at Refs.
\cite{eiqcm,eiecaja,eildiquark}, we only briefly review it here.  The
meson-nucleon (in general the meson-baryon) interaction vertex is
within the quark model represented by the Feynman diagram shown in
Fig.  1.  The quark-hadron vertex is in the quark model described by
the interaction Lagrangians of the form

\begin{equation}
{\cal L}_H(x)=g_H \; H(x) \; J_H(x) \;,
\end{equation}
     where $J_H(x)$ are quark currents with the quantum numbers
corresponding to the considered hadronic field $H(x)$.  The
renormalized coupling constant $g_H$ for a hadron of a given mass can
be obtained from the following compositeness condition

\begin{equation}
Z_H =
1 + {{3 g_H^2} \over {(2 \pi )^2}} \; \tilde{ \Pi}'_H (m^2_H) = 0 \;,
\label{composit}
\end{equation}
     where $\tilde{ \Pi}'_H$ is the derivative of the hadronic mass
operator.

     Let us specify the actual Lagrangian for both types of vertices
we have in Fig.  1, i.e., the quark-meson vertex and the quark-baryon
vertices.  The quark-meson interaction Lagrangian reads

\begin{equation}
{\cal L}_M={g_M\over{\sqrt 2}} \; \sum_{i=1}^8 M_i \; \bar q
\: \Gamma_M \; \lambda_i \; q \; ,
\label{mesonlagrangian}
\end{equation}
     where \ $q,\bar q $ are the quark, antiquark meson constituting
fields, $\bar q = (\bar u,\bar d,\bar s)$, $M_i$ are the Euclidean
mesonic fields relating to the physical mesons in the standard way
\cite{eildiquark}, \ $\lambda_i$ are the Gell-Mann matrices, and
$\Gamma_M$ stands instead of \ $i\gamma^5$ for pseudoscalar mesons
$P(\pi,\eta,\eta^\prime)$, \ $\gamma^\mu$ for vector mesons
$V(\rho,\omega,\phi)$, and $(I-iH_S\hat\partial/\Lambda_q)$ for scalar
mesons $S(a_0,f_0,\varepsilon)$.  Because of SU(3) breaking, the
singlet and octet mesons are mixed as follows

\begin{eqnarray}
 M
\to \cos\delta_\Gamma
\left({\bar uu+\bar dd \over \sqrt{2}} \right)
-(\bar ss) \sin\delta_\Gamma,
 \nonumber\\
 M'
\to -\sin\delta_\Gamma\left({\bar uu+\bar dd\over\sqrt{2}} \right)-
(\bar ss)\cos\delta_\Gamma, \\
\label{mixing}
 M  \equiv (\eta^\prime, \omega, \varepsilon); \;
M' \equiv (\eta, \phi, f_0(975)) \nonumber
\end{eqnarray}
     where $\delta_\Gamma=\theta_\Gamma-\theta_{I\Gamma}$ and
$\theta_{I\Gamma}=35^\circ$ is the so-called ideal mixing angle.  The
mixing angles of pseudoscalar and vector mesons are chosen to be equal
to $\delta_P=-46^\circ$ and $\delta_V=0^\circ$, respectively.  The
scalar meson parameters $\delta_S$, $H_S$, and $m_\varepsilon$
are supposed to be free.
Their determination we will comment on in the next subsection.

     The SU(3) quark currents with baryon quantum numbers have to be
symmetric in respect to the quark field permutation.  Since, for the
$(1/2)^+$ baryonic octet there are two independent three-quark
currents, the quark-baryon interaction Lagrangians read

\begin{eqnarray}
{\cal L}_{B}  & = & {\cal L}_{BT} + {\cal L}_{BV} ,  \\
{\cal L}_{BI} & = & g_{BI} \; \bar B \;  J_{BI}      \\
              & = & ig_{BI} \; \bar B_j^k \; R^{kj; \; j_1,j_2,j_3}_I
\; q_{j_1}^{a_1} q_{j_2}^{a_2} q_{j_3}^{a_3}
\; \varepsilon^{a_1a_2a_3} + H.c.           \nonumber
\end{eqnarray}
     In these expressions $j=(\alpha ,m)$; \ and $(a_i,\alpha_h,m_i)$
are the colour, spin, and flavour indices, respectively.  $B_j^k$ \
are the Euclidean baryonic fields, and matrices $ R^{kj;j_1j_2j_3}_I $
provide proper quark content of the baryons in the vector or tensor
coupling scheme, $I = V,T$.

     The meson-nucleon interaction is in a quark model represented
through the diagram as in Fig.  1.  The typical matrix element
corresponding to the process $B\to B+M$ is proportional to the
following expression

\begin{eqnarray}
\int \! d\sigma_{vac} \; \bar B(x_1) \; S(x_1x_3|B_{vac}) \; M(x_3)\
S(x_3x_2|B_{vac}) \; B(x_2) \nonumber \\
\times
\int \! d\sigma_{vac^\prime} \; Tr[S(x_1x_2|B_{vac^\prime}) \;
S(x_2x_1|B_{vac^\prime})]\ ,
\label{twoloop}
\end{eqnarray}
where $S(x,x'\vert B_{vac})$ denotes the quark propagator in the
external gluon field $B_{vac}$ and $d\sigma_{vac}$ is the measure of
integration over $B_{vac}$.  This highly complex gluon vacuum is
supposed to provide quark confinement itself within QCD.


     To proceed in evaluation of the expression (\ref{twoloop}) we
make use of the QCM method.  The cornerstone of this effective field
theory is a prescription for parameterization of the confinement
producing gluon vacuum fields \cite{eiqcm,eiecaja}.  This means that
the expression (\ref{twoloop}) is substituted by the following one:

\begin{eqnarray}
\int \! d\sigma_v \; \bar B(x_1) \; S_v(x_1-x_3)
   \;  M(x_3) \; S_v(x_3-x_2)  \; B(x_2)
\nonumber \\
\times
\int \! d\sigma_{v^\prime} \; Tr[S_{v^\prime}(x_1-x_2) \;
   S_{v^\prime}(x_2-x_1)] \;,
\label{ansatz}
\end{eqnarray}
which is the QCM ansatz \cite{eiqcm,eiecaja}.   In this expression

\begin{eqnarray}
S_v(x_1-x_2)=\int {d^4p\over (2\pi)^4i}  \;
{e^{-ip(x_1-x_2)} \over v\Lambda_q - \hat p}
\label{qpropagator}
\end{eqnarray}
     is the quark field propagator weighted by the quark confining
field parameter $v$.  The model parameter $\Lambda_q$ determine the
confinement range.  The indefinite measure $d\sigma_v$ in
(\ref{ansatz}) is defined as

\begin{equation}
\int{d\sigma_v\over v-z}=G(z)=a(-z^2)+zb(-z^2) \; .
\label{confinementfunction}
\end{equation}
     The function $G(z)$, the so-called confinement function, is the
entire analytical function that decreases faster than any degree of z
in the Euclidean direction $z^2\to -\infty$.  This requirement gives
us a possibility to construct the finite theory with confined quarks.
Note that this requirement is very general and as a result we can
choose the various actual forms of $G(z)$.  The confinement function
is taken here to be universal, i.e., it is colour and flavour
independent, and unique for all quark diagrams determining hadron
interactions.  As an experience has shown, the only its integral
characteristics are important for description of the low energy
physics.

     To simplify the calculations of the Feynman diagram in Fig.  1,
we can substitute the inner two-quark loop by the single propagator,
the so-called diquark propagator, Refs.  \cite{eildiquark}.  The
meson-baryon vertex of Fig.  1 will then be redrawn to that one shown
in Fig.  2.  This means that the subdiagram corresponding to the
independent two-quark loop

\begin{eqnarray}
\Pi^{\Gamma_1\Gamma_2}(p)=
\int{d^4k\over 4\pi^2i}\int d\sigma_{v^\prime} \;
Tr[\Gamma_1^\prime \; S_{v^\prime}(p+k) \; \Gamma_2^\prime \;
S_{v^\prime}(k)]
\label{Smtxdiquark}
\end{eqnarray}
     is substituted by the diquark propagator $D^{\Gamma_1\Gamma_2}$

\begin{eqnarray}
D^{\Gamma_1\Gamma_2}(k)={d^{\Gamma_1\Gamma_2}\over{M^2_D - k^2}} \;,
\label{diquarkpropagator}
\end{eqnarray}
     where $M_D$ is a diquark mass and $d^{\Gamma_1\Gamma_2}$ are
coefficients dictated by the symmetry properties.  $d^{VT} = -d^{TV} =
(ik_\alpha g^{\mu\beta}-ik_\beta g^{\mu\alpha})$.  This approximation
should fulfill the general requirement - not to break the relation
between the baryon electromagnetic vertex and the mass operator, the
Ward identity.  This identity with the compositeness condition
(\ref{composit}) give us needed symmetry properties, see Refs.
\cite{eildiquark}.
     Consider the last approximation, the meson-baryon vertex may be
written in the form

\begin{eqnarray}
\Lambda_{MNN} (p,p^\prime)  =  \int {d^4k\over \pi^2i}
                                 \int  d\sigma_{v}
\Gamma_1    {1\over {v\Lambda_q-(\hat k -\hat q)}}
  \Gamma_M    {1\over {v\Lambda_q-\hat k}}
            {d^{\Gamma_1\Gamma_2} \over {M_D^2-(p-k)^2}}
  \Gamma_2 .
\label{MBBvertex}
\end{eqnarray}
     Finally, the transferred momentum ($q = p - p'$) dependent
meson-nucleon coupling constants are related to this vertex function
as

\begin{equation}
\Lambda_{MNN} (p,p') \; = \; T_M \; G_{MNN} (q^2) \; ,
\end{equation}
     where $T_\pi = \vec\tau i\gamma^5$ , $T_\eta = T_\eta' = I
i\gamma^5$ , $T_{a_0} = \vec\tau$, $T_\varepsilon = T_{f_0} = I$ , for
pseudoscalar and scalar mesons, respectively, and in terms of the
vector and tensor form factors

\begin{eqnarray}
\Lambda_{MNN}^\mu(p,p')\;=
T_M \left[ \gamma^\mu G_{MNN}(q^2)
-i \sigma^{\mu\nu} q_\nu F_{MNN}(q^2) \right] \; ,
\end{eqnarray}
     where $T_\rho=\vec\tau$, $T_\omega = T_\phi = I$ for vector
mesons.  The vector and tensor meson-nucleon coupling constants are
the $G_{MNN}(q^2)$ and $F_{MNN}(q^2)$ taken at the zero transferred
momentum.

\subsection{  QCM parameterization      }
\noindent
     Free parameters of the present QCM version are the parameters of
the quark-meson interaction Lagrangian, the confinement ansatz
parameters, and the diquark propagator parameter.

     As stated above the quark-meson interaction Lagrangian
(\ref{mesonlagrangian}) has its free parameters only in the scalar
meson sector.  These are the derivative term strength $H_S$, and the
mixing angle value $\delta_S$.  Both parameters have been determined
and thoroughly discussed in Ref.  \cite{eiecaja}.

     The confinement ansatz (\ref{ansatz}-\ref{confinementfunction})
free parameters are the coefficients of the confinement functions
$a(u)$, $b(u)$ and the light quark confinement parameter $\Lambda_q$.
As follows from eqs.  (\ref{ansatz}, \ref{confinementfunction}), the
confinement functions $a(u)$ and $b(u)$ should be entire analytical
functions decreasing sufficiently rapidly in the Euclidean region
$Re(u)\to \infty$.  In this paper we take these functions in the
simplest forms

\begin{eqnarray}
& &a(u)=a_0 \; exp(-u^2-a_1u) \;, \\
& &b(u)=b_0 \; exp(-u^2+b_1u) \;.
\end{eqnarray}
     The coefficients ($a_0, a_1, b_0, b_1$), the light quark
confinement parameter $\Lambda_q$, and the diquark mass $M_D$ have
been chosen to fit a convenient set of reference observables
\cite{eiltomsk}.  The chosen set of the reference hadronic processes
and the resulting QCM values are shown in Table \ref{reference}
together with the reference empirical data.  These results have been
obtained with the QCM parameters, which are shown in Table
\ref{qcmpar}.

     The obtained parameter set have been used to predict numerous
characteristics of hadrons and hadronic processes as magnetic moments
of baryons, weak coupling constants, and various decay widths with
rather good results in Ref.  \cite{eiltomsk}.  That work is in fact a
reinvestigation of the same physics as has been studied in the paper
\cite{eildiquark} using but different -- the constant-mass form of
the diquark propagator.  The application of such form of the diquark
propagator has been motivated by the success of the bottom mesons
decay studies, where for a heavy-quark propagator the constant-mass
propagator has been used.

     As one can see in Refs.  \cite{eiecaja,eildiquark,qcm} generally
good assessments of the strong, electromagnetic, and weak interactions
controlled processes have been acquired within the QCM.  It should be
said further that after preceding steps we have the effective quark
field theory without {\em free parameters}.  Within this frame we can
calculate the meson-nucleon coupling constants without additional free
parameters too.

     \section{ Results and Discussion }
     \subsection{ Introduction }
\noindent
     Despite of the fact that the one-boson exchange model is a
simplified representation of the NN forces, the effectiveness of this
approach is at least at low energies established, see, e.g.  Ref.
\cite{goh}.  The strong intermediate range attraction and the strong
short range repulsion bring, however, some questions concerning their
microscopic understanding.  Within OBE models are these NN force
properties described by using the $\varepsilon$ and $\omega$ mesons,
to describe the attractive and repulsive forces, respectively
\cite{bj,machleidt}.

     Many studies have been devoted to elaborate understanding of the
intermediate range attraction.  The studies performed with dispersion
relation techniques conclude \cite{durso} that a major part of this
attraction arises from the correlated two-pion exchanges, which are in
turn well approximated by the exchange of the $\varepsilon$ meson.
The similar results have been obtained in the field-theoretical
approach of Ref.  \cite{plomon} and by Bonn group \cite{mhe} within
their full meson exchange model.

     There are also other papers, which have studied the intermediate
range attraction in a more microscopic way.  Thus, using the soliton
model Kaiser and Meissner have shown that the inclusion of the pion
loops gives the intermediate range attraction with the strength
compared to the Paris potential \cite{2pichsm}.  The new information
arises also from a development by Weinberg \cite{weinberg} and others
\cite{ordonez}, who have recently use a chiral perturbation theory to
study the nature of the NN forces.  A satisfying qualitative feature,
which they have found shows that the uncorrelated two-pion exchange
with some of the higher order contact terms provide the intermediate
range attraction too.  For discussion of the $\varepsilon$ meson
itself see also very recent development by the Brooklyn group
\cite{brooklyn}.  Other treatments of low-lying scalar mesons can be
find also in \cite{lanik}.

     It was known for a long time that the short range repulsive force
produced by the $\omega$ meson exchange partially simulates forces
originating from the quark and gluon exchange processes and from
heavier vector and tesor meson exchanges also.  The understanding of
these processes advanced recently also.  A qualitative understanding
of the short range part of the NN forces has been obtained as produced
by the one-gluon exchange in the resonating-group method based quark
model \cite{okay}, or as a ``van der Waals' repulsion" in the chiral
bag language \cite{vento}.  In a more refined model version
\cite{shimizu} not only gluons but also Goldstone pions are
exchanged between quarks.  Their exchange immediately followed by a
quark pair exchange add further short range repulsion revealing thus
abundance of structures behind the effective OBEP's $\omega$ exchange.
As have been discussed \cite{myhrer} such calculations can easily
accommodate a repulsive $\omega$ exchange using the $\omega$NN
coupling compatible with the SU(3) value.

     One can see from this discussion that within OBE models we use at
least two-mesons, the exchange of which simulate the more complex
exchanges too.  Consequently, the couplings of these effective
exchange fields differ from the couplings of the elementary processes
and also from that ones calculated within the QCD structure level
based models.  Bearing this in mind, we can go on to discuss our
results.

     \subsection{ Included Meson Exchange Fields}
\noindent
     The empirical findings show that to describe the low energy NN
scattering the pseudoscalar, vector, and scalar meson fields are
necessary to generate the exchange forces \cite{bj,machleidt}.  To be
consistent within the quark model framework we should consider the
mesons constructed from the $u, d, s$ quarks.  Thus, in the present
paper, we take as a set of exchanged mesonic fields the SU(3)
pseudoscalar, vector, and scalar mesons.  Accordingly, our set of the
mesonic fields consists of the $\pi, \eta, \eta'$ pseudoscalars, the
$\rho, \phi, \omega$ vectors, and the $a_0, f_0(1400)$ scalar meson
fields.  Furthermore, we consider the scalar-isoscalar
$\varepsilon(600)$ meson also.  The $f_0(975)$ meson we do not include
in this work because of its coupling constant.  The present
parameterization of the QCM predicts it to be very small, of about
0.2.  The $\eta'$ meson we should include to be consistent from the
quark model point of view, where $\eta$ and $\eta'$ are formed in
pseudoscalar octet-singlet mixing \cite{pdg}.  The QCM predictions of
the meson-nucleon coupling constants, are shown in Table \ref{obep}.
Some of these coupling constants are connected by the SU(3) symmetry
relations.  These are, the vector couplings for the $\rho$ and
$\omega$ vector mesons.  Further, the ratio of the vector meson Pauli
to Dirac coupling constant are $\kappa_{\rho} = f_{\rho NN} / g_{\rho
NN} = (\mu_p-1) - \mu_n$ and $\kappa_{\omega} = f_{\omega NN} /
g_{\omega NN} = (\mu_p-1) + \mu_n$.

     \subsection{ OBEP Construction}
\noindent
     The construction of our OBE QRBA9 (Quark Relativistic Bosons
version A with 9 exchanged fields) model we start out from our QCM
predictions of the meson-nucleon coupling constants and typical cutoff
masses \cite{machleidt} of the phenomenological form factors
(\ref{ff}).  Because of the above specified reasons, which are
connected with the effectiveness of $\varepsilon$ and $\omega$
exchanges, we first optimize the $\varepsilon$NN and $\omega$NN
coupling constants.  Afterward, we include to optimization process
also the form factor cutoff masses.  Parameters of the resulting QRBA9
OBE model are also shown in Table \ref{obep}.

     Concerning the parameter determination procedure the following
should be said.  As the empirical data we take the phase shift values
obtained by Arndt and collaborators \cite{arndtnn}.
The fitting we span over up to 450 MeV of the nucleon
laboratory energy.  From a physical viewpoint this may be done because
the imaginary parts of the phase shifts in all partial waves are
small, except the $^1D_2$ and $^3F_3$ waves, in this energy region
\cite{bj}.

     \subsection{ Phase Shifts}
\noindent
     Our phase shifts are shown in Figs. 3-4.  They are compared there
to empirical data and to a phenomenological fit.  The referred
empirical data are that ones, which we use in our fitting procedure
\cite{arndtnn}.  The results of the mentioned phenomenological fit we
calculate from the Bonn OBEP(B) model \cite{machleidt}, commonly
regarded as a standard one.  Note that this OBE model is affirmed to
325 MeV of the laboratory energy.

     As seen, our predictions agree well with the empirical data.  To
quantify this statement we can say that the ratio of the $\chi^2$
criteria, which we obtain with the QRBA9 model, to that one, we obtain
with the Bonn OBEP(B) model \cite{machleidt}, is 1.23.  Note that
because of the coupling to the isobar channel, which is, as it has
been shown by Lomon \cite{lomon}, responsible for the resonant
behaviour of the $^1D_2$ and $^3F_3$ phase shifts, we do not show
these phase shifts here (taking them into account the mentioned
$\chi^2$ ratio will be 1.80).

     Regarding the phase shifts we would like to comment on the
behaviour of the $^3S_1-^3D_1$ mixing parameter $\varepsilon_1$ only.
In a recent phase shift analysis, in which the Basel group has used
their newly measured spin correlation parameter in a neutron-proton
scattering, they have obtained the value of $\varepsilon_1 = 2.9^0
\pm0.3^0$ at 50 MeV \cite{basel}.  The another analysis \cite{KSS92},
which includes also the Basel data, reports the value of
$\varepsilon_1 = 2.2^0 \pm0.5^0$ for the same energy.  Our prediction
is just below 2.6$^0$, which is consistent with both mentioned
analyses.

     To realize the quality of the present phase shifts description,
it should be articulated that this should be compared rather to other
existing (semi)microscopic results than to the fully phenomenological
fit as the Bonn model is.  Thus, Fujiwara and Hecht, after their
impressive development of the nonrelativistic resonating group method
based model, have concluded that they have obtained the
semiquantitative fit of the NN phase shifts \cite{fh}.  Similar
results as that ones have been obtained also in Ref.  \cite{hofmann}.
After a development described in Refs.  \cite{shimizu}, a
semiquantitative fit of the NN scattering data have been obtained in
Ref.  \cite{faesslernn}, using but two different couplings for the
isoscalar-scalar meson-nucleon vertex.

     \subsection{ Comparison of QCM Predictions with Nonlinear Chiral
Effective Lagrangian Predictions and OBEP Model Parameters }
\noindent
     Now we compare the QCM coupling constant predictions with that
ones obtained within the framework of a nonlinear chiral meson theory
in which nucleons emerge as topological solitons (ChSM).  Notice that
even though accentuating different aspects, both methods, QCM and
ChSM, have been deduced from QCD.  The ChSM coupling costants have
been calculated in two model versions in Ref.  \cite{chsm} with
inclusion of the $\pi$, $\rho$, and $\omega$ mesons as explicit
degrees of freedom.  The complete model accounts not only for the
chiral anomaly of the underlying QCD through the Wess-Zumino term
governing the $\omega$ meson coupling to the topological baryon
current, as minimal model does, but has also that part of the
Wess-Zimino action, which incorporates additional $\pi\rho\omega$
couplings.

     Further, we compare theoretical predictions to the meson-nucleon
coupling constants from other NN scattering studies.  The QCM and both
ChSM predictions are together with empirical coupling constants that
have been obtained by two leading groups in the low energy NN
scattering phenomenology, namely the Bonn and Nijmegen groups, shown
in Table \ref{compar}.

     \subsubsection{ Pseudoscalar meson-nucleon couplings}
\noindent
     The inspection in Table \ref{compar} reveals that all QCM
coupling constants are lower than ChSM predictions.  As seen from the
difference between the minimal and complete ChSM models, the mentioned
$\pi\rho\omega$ coupling term decrease the $\pi$NN coupling by 10 \%,
bringing it closer to our prediction.  Nevertheless, the complete
model prediction is still 13 \% higher.  Note that the much closer
result (14.44) have been obtained by Ho$\!\!\!/$gaasen and Myhrer
\cite{cbmpinn} within but the cloudy bag model.

     A present warm discussion \cite{ericson,nijpin,machlli} about the
$\pi$NN coupling constant has appeared on account of the analysis of
new $\pi$N scattering data by Arndt and coworkers \cite{arndtpin}.  In
that work the VPI\&SU group has estimated the charged-pion coupling
constant to be 13.31 $\pm$0.27 being thus in the middle between
earlier results of the Nijmegen group.  The last group has found first
$g^2/4\pi$ = 13.11 $\pm$0.11 in \cite{nijpin87} and later $g^2/4\pi$ =
13.55 $\pm$0.13 in \cite{nijpin90}.  Problems arise because these new
values of the $\pi$NN coupling are much lower than the coupling
constant value commonly used for more than a decade, namely 14.28
$\pm$0.18, see, Refs.  \cite{koch,machleidt}.  The primary instruction
coming from this development is that the stated errors, being
statistical only on the level of 1\%, seriously underestimate the real
uncertainty \cite{ericson,machls}.  In the last reference it has been
shown as well that the minimal value of the $g^2_{\pi}/4\pi$, which is
necessary to describe the deuteron quadrupole moment is 13.65 for the
Bonn OBEP \cite{machleidt}.  The QCM predicts the $\pi$NN coupling
constant, which is close to the last value and to the Nijmegen value.

     The QCM prediction of the $\eta$ meson-nucleon coupling is near
to the Nijmegen coupling and between the Bonn values.  The Nijmegen
$\eta'$NN coupling is close to our prediction too.

     \subsubsection{ Scalar meson-nucleon couplings}
\noindent
     Comparing the QCM predicted $a_0$NN coupling constant to
phenomenological values obtained in the NN scattering fit, we find
that our value is between the Bonn$^M$ model and the Nijmegen model
values.  All together are but much higher than the Bonn$^H$ value.

     Since we do not know other predictions of the $f_{0}(1400)$NN
coupling constant we do not show this value in Table \ref{compar}.
Note, however, that a decrease of its coupling in comparison with the
$\varepsilon$NN coupling shows the mass dependence of the
scalar-isoscalar meson-nucleon coupling.

     \subsubsection{ Effective scalar and vector meson-nucleon
couplings}
\noindent
     As we mentioned already, the intermediate range attractions have
been obtained from the correlated two-pion exchange, which has been
evaluated with the minimal model Lagrangian of ChSM in Ref.
\cite{2pichsm}.  The resulting two-pion strength between solitons then
authors have parameterized by exchange of one-scalar-isoscalar meson
with the shown coupling constant.  The range of uncertainty shown
there originate in different ad hoc taken pion-loops renormalizing
cutoffs.  The authors claim, however, that taking into account the
full Lagrangian coupling should lead to a higher predicted
$\varepsilon$NN coupling and thus perhaps close to the QCM result.

     Within the OBE models we use the effective scalar-isoscalar meson
as the intermediate range attraction simulating field.  Consequently,
the constraint on its coupling constant requires of it to has a
reasonable value and to be in a strong correlation with the value of
the short range repulsive strength, which is generated by the
effective $\omega$ vector meson-nucleon coupling.  Thus, as we explain
already, these effective fields simulate the more complex exchanges as
well and therefore the values of their couplings cannot be compared to
the QCM nor ChSM results.

     \subsubsection{ Vector meson-nucleon couplings}
\noindent
     {\em Dirac coupling.} The accepted $\rho$NN vector couplings are
about 0.43 $\pm$0.10 \cite{bj}, 0.5 \cite{bbn}, which agree with the
QCM prediction.  These values are compatible with the minimal model
prediction but the $\pi\rho\omega$ coupling term of the complete model
increase it by additional 47 \%.  So, the ChSM predictions are 1.36
and 2 times higher then our result.

     Our value of the Dirac $\rho$NN coupling is close to the Bonn$^H$
value, which is but about a factor of 2 lower than Bonn$^M$ one.  Note
that the Bonn$^H$ OBEP is the same for both isospin channels, like the
QRBA9 one, and unlike the Bonn$^M$ set.  The Nijmegen OBEP $\rho$NN
coupling is too far from accepted values also, but together with
Bonn$^M$ value agree with the prediction of the complete ChSM version.

     As QCM is constructed within the SU(3) frame, the QCM ratio of
the $\omega$ to $\rho$ vector meson couplings is also SU(3) one,
namely $g^2_{\omega} / g^2_{\rho}$ = 9.  Thus the SU(3) expected
$\omega$NN vector coupling is about 4.  Both ChSM predictions
dynamically violate this symmetry being 10.8 and 7.4 for the minimal
and complete models, respectively.  Nevertheless, ChSM predictions are
1.64 times higher of our value, which is just in the middle between
rates we have found for the $\rho$NN couplings.

     {\em Pauli coupling.} It is interesting to observe that the ChSM
Pauli couplings of the vector mesons have been shifted more closely to
the QCM predicted values when $\pi\rho\omega$ coupling term is
included also.  Thus, the complete model Pauli to Dirac ratio
predictions are much close to the empirical findings.

     The size of the Pauli coupling constant, which is in common use
in OBE models have relied \cite{machleidt} on the old analysis of Ref.
\cite{hohler}.  In that paper the value of $\kappa_{\rho}$ = 6.1
$\pm$0.6 was obtained for a ratio of the Pauli to Dirac coupling
constants.  It is known but that the $\kappa_{\rho}$ has to be
consistent with the empirical value following from a vector meson
dominance of a low momentum part of the nucleon electromagnetic form
factor.  The empirical value of $\kappa_{\rho}$ = 3.7 \cite{bj} is
consistent with our QCM value $\kappa_{\rho}$ = 3.66 and with the
Nijmegen model value $\kappa_{\rho}$ = 4.221.

     A recent study of the deutron properties by Machleidt and
Sammarruca \cite{machls} shown that to use $\kappa_{\rho}$ = 3.7 in
the Bonn$^M$ OBEP $g^2_{\pi}/4\pi$ = 13.65 is needed.  This finding
is, however, close to the QCM predictions.

     \subsection{ Form Factors}
\noindent
     Expected form factor cutoff masses for different elementary
meson-nucleon vertices would not be very different from the well known
value of the electromagnetic cutoff mass.  The last value is well
described in QCM as in ChSM \cite{chsmelmax} as well as in lattice QCD
calculations \cite{lattelm}, and quark bag models
\cite{myhrer}.

     The only experimentally studied meson-nucleon vertex is the
$\pi$NN one.  For this vertex, model calculations as well as empirical
analyses favour the $\Lambda_{\pi}$ value of about 0.9 GeV.  On the
other hand, a more indirect process as the NN scattering is, require a
much higher value.  It thus appears that $\Lambda_{\pi NN}$ needed in
the NN scattering is an effective quantity.

     As we have mentioned in sect.  1, the expected medium vacuum
fluctuations change meson and nucleon propagations \cite{lurie} and
create a variety of different forms of correlations on both structure
levels, the QCD as well as the QHD levels.  Second, as known, and as
it has been recently calculated in Refs.  \cite{meissner,qiansusu},
the meson-nucleon interaction is density dependent.  Whether this
density dependence of a bare meson-nucleon system is effective enough
to modify the low energy NN scattering is a priori not known.
Further, it is known that the couplings depend also on a type of
relativistic equation in use (to observe this one can intercompare
OBEP parameters shown in Tables A.1 and A.2 of the Ref.
\cite{machleidt}) and on a chosen spectrum of exchanged particles
(compare, e.g., different OBE models of the Ref.  \cite{goh}).
Therefore, in the present work we do not intend to solve this form
factor problem, but we parameterize it, as all others do also.

     Within the present environment, composed of the
Blankenbecler-Sugar-Logunov-Tavkheli-dze equation and the QRBA9 OBE
model, most of used cutoffs are not very certainly determined.  The
cutoff masses of the $\eta$, $\eta'$ and $a_0$ mesons may be changed
in a wide interval of values without a significant deterioration of
the fit quality.  The $\pi$NN and $\rho$NN vertex cutoffs are on the
other hand determined strictly with their correlation measuring
-97$\%$.

     Although, we do not apply the QCM predicted cutoffs here, it is
of specific interest to compare them with other findings, especially
for the critical $\pi$NN and $\rho$NN vertices.  The QCM predicts for
the $\Lambda_{\pi}$ and $\Lambda_{\rho}$, the values of 0.88 and 0.60
GeV, respectively.  These predictions may be compared to the results
of a non-linear chiral meson theory of Ref.  \cite{chsm}, where the
values obtained for $\Lambda_{\pi}$ and $\Lambda_{\rho}$ are 0.86 and
0.93 GeV, respectively.  Although the absolute values of the QCM are
very different from the QRBA9 model values, Table \ref{obep}, we find
that their ratio is 1.477 comparing to the QRBA9 value of 1.494.  The
mentioned density dependence of the form factor cutoffs, estimated in
\cite{meissner} for $\Lambda_{\pi}$ and $\Lambda_{\rho}$, represents
approximately 20 and 10$\%$ reduction at the nuclear density,
respectively.  On the other hand, a comparison of the QRBA9 cutoffs
with the QCM ones shows that the former values are changed by a factor
higher than two.  This indicate that the density dependence of the
form factor cutoffs may be neglected here.

     The high $\Lambda_{\pi NN}$ value seems to be nevertheless
understandable as an effect produced by vacuum fluctuations in meson
exchange currents studied as a renormalization problem in meson
theories \cite{lurie}.  As recently has been suggested by Ueda
\cite{ueda} accounts for the role of the $\pi\rho$ and
$\pi\varepsilon$ loops explain satisfactorily a difference between the
$\Lambda_{\pi NN}$ and $\Lambda_{\pi,(3-body)}$ values.  Here, the
last $\Lambda_{\pi,(3-body)}$ cutoff, which has a value consistent
with our $\Lambda_{\pi}$ value, is the pion-nucleon cutoff one needs
to interpret a nuclear three-body problem.

     \section { Conclusion }
\noindent
     In this paper we have investigated the low energy part of the NN
scattering relying on the empirical findings that the mesonic degrees
of freedom represent a proper way to describe hadron scattering at
this energy domain \cite{bj}.  In this, we have constructed the one
boson exchange model using the meson-nucleon couplings predicted by
QCM, a model deduced from the QCD.  It is to be mentioned that it was
not our intention to find a quantitatively competitive description of
the NN scattering observables in the present work.  Rather, we intend
to find a way to understood the NN scattering (generally baryon-baryon
scattering) observables having a model parameterized on the quark
level only, although it may not be achieved early, and a some guidance
should be gathered step by step.

     Thus the present NN scattering description is composed of the
Blankenbecler-Sugar-Logunov-Tavkhelidze quasipotential equation and
the QRBA9 OBE model.  Constructing the QRBA9 OBEP we take the
meson-nucleon copuling constants as we obtain them as parameter-free
predictions of the QCM.  The intermediate range attraction and the
repulsive short range components of the NN forces we describe as usual
through the effective $\varepsilon$ and $\omega$ exchanges.  The
cutoff masses we subjugate to fit the empirical phase shifts as common
in all models constructed up to now.  In result we find that our phase
shifts well agree with the empirical data.  The ratio of our $\chi^2$
which we obtain with the QRBA9 model to the $\chi^2$ we calculate for
the Bonn model \cite{machleidt} is 1.23, what is the unexpectedly good
result.

     The $\Lambda_{\pi NN}$ and $\Lambda_{\rho NN}$ cutoffs we obtain
here are compatible with common OBEP values, however, they are much
higher than that ones predicted by the QCM.  This observation indicate
that the present $\pi$ and $\rho$ meson exchanges effectively simulate
some other processes too.  Ueda's explanation of this problem with
using vacuum fluctuations in the $\pi$ exchange channel suggests that
the same mechanism may be capable to account for cutoff mass
differences in $\rho$ meson exchange channel too.

\section{Acknowledgements}
\noindent
     We have started this study while working at the LTP JINR, Dubna,
Russia.  The authors are indebted to G.V.  Efimov, M.A.  Ivanov and J.
L\'anik for discussions.

\newpage
\begin{center}
{\bf Table 1.}\\
{\em Reference hadronic processes.  Experimental data
taken from \cite{pdg}.}
\begin{tabular}{l l l r}\\
\hline
Process & Observable value &Experiment & QCM \\
\hline
$\pi\to\mu\nu$ & $f_\pi \;$(GeV) & 0.132 & 0.131 \\
$\rho\to\gamma$ & $g_{\rho\gamma}$ & 0.20 & 0.18 \\
$\pi^0\to\gamma\gamma$ & $g_{\pi\gamma\gamma}\;$(GeV$^{-1}$) & 0.276
& 0.287 \\
$\omega\to\pi\gamma$ & $g_{\omega\pi\gamma}  \; $(GeV$^{-1}$) & 2.54
& 2.02 \\
$\rho\to\pi\pi$ & $g_{\rho\pi\pi}$ & 6.1 & 6.5 \\
$p\to p\gamma$ & $\mu_p$ & 2.793 & 2.798 \\
$n\to n\gamma$ & $\mu_n$ & -1.913 & -1.864 \\
\hline\\
\label{reference}
\end{tabular}
\end{center}

\begin{center}
{\bf Table 2.}\\
{\em QCM parameters. The $\Lambda_q$, $M_D$, and $m_\varepsilon$
are in MeV and $\delta$ in degrees.}
\begin{tabular}{l l l l l l l l l}\\
\hline
$a_0$ & $b_0$ & $a_1$ & $b_1$ & $\Lambda_q$   & $M_D$    &
$H_S$ & $\delta_S$ & $m_\varepsilon$                    \\
\hline
1.8 & 2.0 & 0.6 & 0.2 & 400 & 670 &0.55  & 17 & 600      \\
\hline\\
\label{qcmpar}
\end{tabular}
\end{center}

\begin{center}
{\bf Table 3.}\\
{\em The QCM predictions of the meson-nucleon coupling constants,
  and the QRBA9 OBEP parameters. Numbers in {\bf bold face} were
  varied   during the fitting procedure. }

\begin{tabular}{ l c c r}\\
\hline
Vertex         &QCM           &\multicolumn{2}{c}{QBRA9  }\\
          &$g^2/4\pi (f/g)$ &$g^2/4\pi (f/g)$  &$\Lambda$ (MeV)\\
\hline
$\pi         $NN  & 13.85        & 13.85        & {\bf 2110} \\

$\eta        $NN  & 3.858        & 3.858        & {\bf 1000} \\

$\eta'       $NN  & 3.065        & 3.065        & {\bf 1000} \\

$\rho        $NN  & 0.416 (3.66) & 0.416 (3.66) & {\bf 1410} \\

$\phi        $NN  & 1.872        & 1.872        & {\bf 1410} \\

$\omega      $NN  & 3.740(-0.07) & {\bf 15.54 (0.0)} & 2000  \\

$\varepsilon $NN  & 3.620        & {\bf 10.46}  &      2000  \\

$a_0         $NN  & 1.996        & 1.996        & {\bf 1800} \\

$f_0(1400)   $NN  & 2.062        & 2.062        & {\bf 2000} \\
\hline\\
\label{obep}
\end{tabular}
\end{center}

\newpage
\begin{center}
{\bf Table 4.}\\
{\em Comparison of the QCM with Chiral Soliton Model predictions
\cite{chsm} of meson-nucleon coupling constants and the Bonn$^M$
OBEP(B) version \cite{machleidt}, the Bonn$^H$ \cite{holinde}, and the
Nijmegen \cite{nij78} couplings.  $^a$--the correlated two-pion
exchange between two solitons has been simulated by the $\varepsilon$
exchange \cite{2pichsm}. }\\
\begin{tabular}{ c c c c c c c }\\
\hline\\
Vertex
& QCM
&\multicolumn{2}{c}{Chiral Soliton Model}
& Bonn$^M$
& Bonn$^H$
& Nijmegen \\
  & &  Minimal       & Complete      \\
\vspace{\baselineskip}
& $g^2/4\pi$
& $g^2/4\pi$
& $g^2/4\pi$
& $g^2/4\pi$
& $g^2/4\pi$
& $g^2/4\pi$  \\
& $(f/g)$
& $(f/g)$
& $(f/g)$
& $(f/g)$
& $(f/g)$
& $(f/g)$   \\
\hline\\

$\pi    NN$
&  13.85
& 17.3          & 15.7
&  14.4
&  14.4
&  13.676              \\

$\eta   NN$
&  3.858
&  ---     &  ---
&  3.0
&  4.9978
&  3.433              \\

$\eta'  NN$
&  3.065
&   ---     &  ---
&   ---
&   ---
&  3.759         \\

$a_0   NN$
&  1.996
&  ---     &  ---
&  2.488
&  0.373
&  1.632          \\

$\varepsilon NN$
&  3.620
&  1.4 - 2.3 $^a$  & ---
&  ---
&  ---
&  ---                        \\

$\omega NN$
&  3.740
&  6.140          & 6.140
&  ---
&  ---
&  ---                         \\

$\omega NN$
&        (-0.07)
&        (-0.21)  &       (-0.07)
&  ---
&  ---
&  ---                         \\

$\rho   NN$
&  0.416
& 0.567         & 0.835
&  0.9
&  0.470
&  0.795                        \\

$\rho   NN$
&         (3.66)
&       (5.38)  &       (4.36)
&         (6.1 )
&         (6.6 )
&         (4.221)               \\

$\phi NN$
&  1.872
&  ---    &  ---
&  ---
&  5.361
&  0.099 \\
\hline
\label{compar}
\end{tabular}
\end{center}

\listoffigures

\noindent
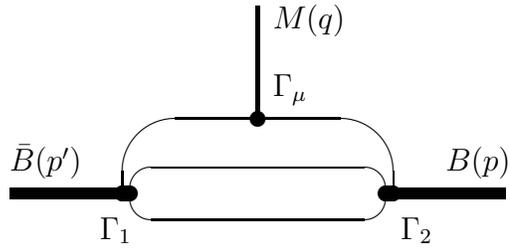
{\bf Fig. 1}
{ The meson -- three-quark-baryon vertex diagram.}\\

\noindent
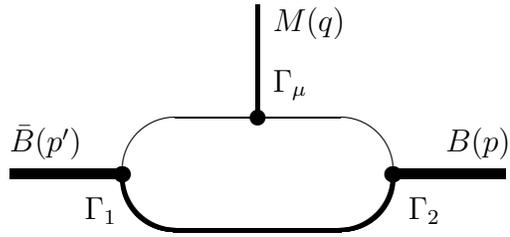
{\bf Fig. 2}
{ The meson -- quark-diquark-baryon vertex diagram.}\\

\noindent
{\bf Fig. 3}
{ The phase shifts of the NN scattering.  The solid lines
represent the results obtained with the QRBA9 parameters.  The dashed
lines refer to the results obtained with the Bonn OBEP(B) model
\cite{machleidt}.
     The circles denote the empirical data of Arndt {\it et al}., Ref.
\cite{arndtnn}.  The cross and the square in $\varepsilon_1$ are the
empirical data of \cite{basel} and \cite{KSS92}, respectively.}\\

\noindent
{\bf Fig. 4}
{ The phase shifts of the NN scattering.  The notation is
the same as in Fig.3.  }\\
%
%
\setlength{\unitlength}{1cm}                                        %
\begin{center}                                                      %
\begin{picture}(14,8)                                               %
          \put(6.5,2.7)     {\oval(3.6,2.0)[t]}                     %
          \put(6.5,2.7)     {\oval(3.4,0.7)}                        %
          \put(4.70,2.7)    {\circle*{0.2}}                         %
          \put(4.71,2.7)    {\circle*{0.2}}                         %
          \put(4.72,2.7)    {\circle*{0.2}}                         %
          \put(4.73,2.7)    {\circle*{0.2}}                         %
          \put(4.74,2.7)    {\circle*{0.2}}                         %
          \put(4.75,2.7)    {\circle*{0.2}}                         %
          \put(4.76,2.7)    {\circle*{0.2}}                         %
          \put(4.77,2.7)    {\circle*{0.2}}                         %
          \put(4.78,2.7)    {\circle*{0.2}}                         %
          \put(4.79,2.7)    {\circle*{0.2}}                         %
          \put(4.80,2.7)    {\circle*{0.2}}                         %
          \put(8.20,2.7)    {\circle*{0.2}}                         %
          \put(8.21,2.7)    {\circle*{0.2}}                         %
          \put(8.22,2.7)    {\circle*{0.2}}                         %
          \put(8.23,2.7)    {\circle*{0.2}}                         %
          \put(8.24,2.7)    {\circle*{0.2}}                         %
          \put(8.25,2.7)    {\circle*{0.2}}                         %
          \put(8.26,2.7)    {\circle*{0.2}}                         %
          \put(8.27,2.7)    {\circle*{0.2}}                         %
          \put(8.28,2.7)    {\circle*{0.2}}                         %
          \put(8.29,2.7)    {\circle*{0.2}}                         %
          \put(8.30,2.7)    {\circle*{0.2}}                         %
          \put(6.5,3.7)     {\circle*{0.2}}                         %
          \put(4.70,2.64)   {\line(-1,0){1.5}}                      %
          \put(4.70,2.65)   {\line(-1,0){1.5}}                      %
          \put(4.70,2.66)   {\line(-1,0){1.5}}                      %
          \put(4.70,2.67)   {\line(-1,0){1.5}}                      %
          \put(4.70,2.68)   {\line(-1,0){1.5}}                      %
          \put(4.70,2.69)   {\line(-1,0){1.5}}                      %
          \put(4.70,2.70)   {\line(-1,0){1.5}}                      %
          \put(4.70,2.71)   {\line(-1,0){1.5}}                      %
          \put(4.70,2.72)   {\line(-1,0){1.5}}                      %
          \put(4.70,2.73)   {\line(-1,0){1.5}}                      %
          \put(4.70,2.74)   {\line(-1,0){1.5}}                      %
          \put(4.70,2.75)   {\line(-1,0){1.5}}                      %
          \put(4.70,2.76)   {\line(-1,0){1.5}}                      %
          \put(8.3,2.64)    {\line(1,0){1.5}}                       %
          \put(8.3,2.65)    {\line(1,0){1.5}}                       %
          \put(8.3,2.66)    {\line(1,0){1.5}}                       %
          \put(8.3,2.67)    {\line(1,0){1.5}}                       %
          \put(8.3,2.68)    {\line(1,0){1.5}}                       %
          \put(8.3,2.69)    {\line(1,0){1.5}}                       %
          \put(8.3,2.70)    {\line(1,0){1.5}}                       %
          \put(8.3,2.71)    {\line(1,0){1.5}}                       %
          \put(8.3,2.72)    {\line(1,0){1.5}}                       %
          \put(8.3,2.73)    {\line(1,0){1.5}}                       %
          \put(8.3,2.74)    {\line(1,0){1.5}}                       %
          \put(8.3,2.75)    {\line(1,0){1.5}}                       %
          \put(8.3,2.76)    {\line(1,0){1.5}}                       %
          \put(6.48,3.7)    {\line(0,1) {1.5}}                      %
          \put(6.49,3.7)    {\line(0,1) {1.5}}                      %
          \put(6.50,3.7)    {\line(0,1) {1.5}}                      %
          \put(6.51,3.7)    {\line(0,1) {1.5}}                      %
          \put(6.52,3.7)    {\line(0,1) {1.5}}                      %
          \put(6.7,4.0)     {$\Gamma_{\rm \mu}      $}              %
          \put(4.4,2.1)     {$\Gamma_1$}                            %
          \put(8.4,2.1)     {$\Gamma_2$}                            %
          \put(3.2,3.0)     {$\bar B(p')$}                          %
          \put(9.0,3.0)     {$B(p)$}                                %
          \put(6.7,4.9)     {$M(q)$}                                %
          \end{picture}                                             %
          \end{center}                                              %
\begin{center}
                                Fig. 1
\end{center}
%
\setlength{\unitlength}{1cm}                                        %
\begin{center}                                                      %
\begin{picture}(10,8)                                               %
          \put(4.5,2.7)    {\oval(3.6,1.5)}                         %
          \put(4.5,2.7)    {\oval(3.56,1.50)[b]}                    %
          \put(4.5,2.7)    {\oval(3.57,1.50)[b]}                    %
          \put(4.5,2.7)    {\oval(3.58,1.50)[b]}                    %
          \put(4.5,2.7)    {\oval(3.59,1.50)[b]}                    %
          \put(4.5,2.7)    {\oval(3.60,1.50)[b]}                    %
          \put(4.5,2.7)    {\oval(3.61,1.50)[b]}                    %
          \put(4.5,2.7)    {\oval(3.62,1.50)[b]}                    %
          \put(4.5,2.7)    {\oval(3.63,1.50)[b]}                    %
          \put(4.5,2.7)    {\oval(3.64,1.50)[b]}                    %
          \put(4.5,2.7)    {\oval(3.60,1.46)[b]}                    %
          \put(4.5,2.7)    {\oval(3.60,1.47)[b]}                    %
          \put(4.5,2.7)    {\oval(3.60,1.48)[b]}                    %
          \put(4.5,2.7)    {\oval(3.60,1.49)[b]}                    %
          \put(4.5,2.7)    {\oval(3.60,1.50)[b]}                    %
          \put(4.5,2.7)    {\oval(3.60,1.51)[b]}                    %
          \put(4.5,2.7)    {\oval(3.60,1.52)[b]}                    %
          \put(4.5,2.7)    {\oval(3.60,1.53)[b]}                    %
          \put(4.5,2.7)    {\oval(3.60,1.54)[b]}                    %
          \put(4.5,2.7)    {\oval(3.56,1.46)[b]}                    %
          \put(4.5,2.7)    {\oval(3.57,1.47)[b]}                    %
          \put(4.5,2.7)    {\oval(3.58,1.48)[b]}                    %
          \put(4.5,2.7)    {\oval(3.59,1.49)[b]}                    %
          \put(4.5,2.7)    {\oval(3.60,1.50)[b]}                    %
          \put(4.5,2.7)    {\oval(3.61,1.51)[b]}                    %
          \put(4.5,2.7)    {\oval(3.62,1.52)[b]}                    %
          \put(4.5,2.7)    {\oval(3.63,1.53)[b]}                    %
          \put(4.5,2.7)    {\oval(3.64,1.54)[b]}                    %
          \put(2.69,2.7)     {\circle*{0.2}}                        %
          \put(2.70,2.7)     {\circle*{0.2}}                        %
          \put(2.71,2.7)     {\circle*{0.2}}                        %
          \put(6.29,2.7)     {\circle*{0.2}}                        %
          \put(6.30,2.7)     {\circle*{0.2}}                        %
          \put(6.31,2.7)     {\circle*{0.2}}                        %
          \put(4.50,3.45)    {\circle*{0.2}}                        %
          \put(2.7,2.64)    {\line(-1,0){1.5}}                      %
          \put(2.7,2.65)    {\line(-1,0){1.5}}                      %
          \put(2.7,2.66)    {\line(-1,0){1.5}}                      %
          \put(2.7,2.67)    {\line(-1,0){1.5}}                      %
          \put(2.7,2.68)    {\line(-1,0){1.5}}                      %
          \put(2.7,2.69)    {\line(-1,0){1.5}}                      %
          \put(2.7,2.70)    {\line(-1,0){1.5}}                      %
          \put(2.7,2.71)    {\line(-1,0){1.5}}                      %
          \put(2.7,2.72)    {\line(-1,0){1.5}}                      %
          \put(2.7,2.73)    {\line(-1,0){1.5}}                      %
          \put(2.7,2.74)    {\line(-1,0){1.5}}                      %
          \put(2.7,2.75)    {\line(-1,0){1.5}}                      %
          \put(2.7,2.76)    {\line(-1,0){1.5}}                      %
          \put(6.3,2.64)    {\line(1,0){1.5}}                       %
          \put(6.3,2.65)    {\line(1,0){1.5}}                       %
          \put(6.3,2.66)    {\line(1,0){1.5}}                       %
          \put(6.3,2.67)    {\line(1,0){1.5}}                       %
          \put(6.3,2.68)    {\line(1,0){1.5}}                       %
          \put(6.3,2.69)    {\line(1,0){1.5}}                       %
          \put(6.3,2.70)    {\line(1,0){1.5}}                       %
          \put(6.3,2.71)    {\line(1,0){1.5}}                       %
          \put(6.3,2.72)    {\line(1,0){1.5}}                       %
          \put(6.3,2.73)    {\line(1,0){1.5}}                       %
          \put(6.3,2.74)    {\line(1,0){1.5}}                       %
          \put(6.3,2.75)    {\line(1,0){1.5}}                       %
          \put(6.3,2.76)    {\line(1,0){1.5}}                       %
          \put(4.48,3.45)   {\line(0,1) {1.5}}                      %
          \put(4.49,3.45)   {\line(0,1) {1.5}}                      %
          \put(4.50,3.45)   {\line(0,1) {1.5}}                      %
          \put(4.51,3.45)   {\line(0,1) {1.5}}                      %
          \put(4.52,3.45)   {\line(0,1) {1.5}}                      %
          \put(4.7,3.8)     {$\Gamma_{\rm \mu}      $}              %
          \put(2.2,2.1)     {$\Gamma_1$}                            %
          \put(6.5,2.1)     {$\Gamma_2$}                            %
          \put(1.2,3.0)     {$\bar B(p')$}                          %
          \put(7.0,3.0)     {$B(p)$}                                %
          \put(4.7,4.6)     {$M(q)$}                                %
          \end{picture}                                             %
          \end{center}                                              %
\begin{center}
                                Fig. 2
\end{center}
\end{document}